\documentclass[letterpaper]{article} % DO NOT CHANGE THIS
\usepackage[submission]{aaai25}  % DO NOT CHANGE THIS
\usepackage{times}  % DO NOT CHANGE THIS
\usepackage{helvet}  % DO NOT CHANGE THIS
\usepackage{courier}  % DO NOT CHANGE THIS
\usepackage[hyphens]{url}  % DO NOT CHANGE THIS
\usepackage{graphicx} % DO NOT CHANGE THIS
\usepackage{natbib}  % DO NOT CHANGE THIS AND DO NOT ADD ANY OPTIONS TO IT
\usepackage{caption} % DO NOT CHANGE THIS AND DO NOT ADD ANY OPTIONS TO IT
\usepackage{algorithm}
\usepackage{algorithmic}
\usepackage{amsmath}
\usepackage{amssymb}
\usepackage{booktabs}
\usepackage{algorithm}
\usepackage{algorithmic}
\usepackage{marvosym}
\usepackage{newfloat}
\usepackage{listings}
\DeclareCaptionStyle{ruled}{labelfont=normalfont,labelsep=colon,strut=off} % DO NOT CHANGE THIS
\floatstyle{ruled}
\newfloat{listing}{tb}{lst}{}
\floatname{listing}{Listing}
\title{PRAGA: Prototype-aware Graph Adaptive Aggregation for Spatial \\
Multi-modal Omics Analysis}
\author {
    Xinlei Huang\textsuperscript{\rm 1},
    Zhiqi Ma\textsuperscript{\rm 1}, 
    Dian Meng\textsuperscript{\rm 1}, 
    Yanran Liu\textsuperscript{\rm 1}, 
    Shiwei Ruan\textsuperscript{\rm 1}, \\
    Qingqiang Sun\textsuperscript{\rm 1}, 
    Xubin Zheng\textsuperscript{\rm 1, \rm 2, }\thanks{Corresponding author}, 
    Ziyue Qiao\textsuperscript{\rm 1}
}
\affiliations {
    \textsuperscript{\rm 1}School of Computing and Information Technology, Great Bay University, Dongguan, 523000, China\\
    \textsuperscript{\rm 2}Guangdong Provincial Key Laboratory of Mathematical and Neural Dynamical Systems, Dongguan, 523000, China\\
    xbzheng@gbu.edu.cn
}

\usepackage{bibentry}
\begin{document}

\maketitle

\begin{abstract}
Spatial multi-modal omics technology, highlighted by Nature Methods as an advanced biological technique in 2023, plays a critical role in resolving biological regulatory processes with spatial context. Recently, graph neural networks based on K-nearest neighbor (KNN) graphs have gained prominence in spatial multi-modal omics methods due to their ability to model semantic relations between sequencing spots. However, the fixed KNN graph fails to capture the latent semantic relations hidden by the inevitable data perturbations during the biological sequencing process, resulting in the loss of semantic information. In addition, the common lack of spot annotation and class number priors in practice further hinders the optimization of spatial multi-modal omics models. Here, we propose a novel spatial multi-modal omics resolved framework, termed \textbf{PR}ototype-\textbf{A}ware \textbf{G}raph \textbf{A}daptative Aggregation (PRAGA). PRAGA constructs a dynamic graph to capture latent semantic relations and comprehensively integrate spatial information and feature semantics. The learnable graph structure can also denoise perturbations by learning cross-modal knowledge. Moreover, a dynamic prototype contrastive learning is proposed based on the dynamic adaptability of Bayesian Gaussian Mixture Models to optimize the multi-modal omics representations for unknown biological priors. Quantitative and qualitative experiments on simulated and real datasets with 7 competing methods demonstrate the superior performance of PRAGA.
Code is available at \url{https://github.com/Xubin-s-Lab/PRAGA}.

\end{abstract}

\section{Introduction}
Spatially resolved transcriptomics was crown as the Method of the Year by Nature~\cite{xiaowei2021method}. This technology expanded the biological view of the gene expression abundance in single cells to spatial context, deciphering cell types and the heterogeneity in complex tissues.
Recently, spatially resolved multi-modal omics, including transcriptomics, proteomics, and chromatin accessibility, were proposed to aggregate these modalities for comprehensively resolving gene regulation and microenvironment with spatial information in complex tissues~\cite{li2024cross,long2024deciphering}.

The major challenge in spatial multi-modal omics is how to encode omics features from different modalities with corresponding spatial information into a unified latent space.
Existing methods mainly build K-nearest neighbor (KNN) graphs to model the feature correlation with spatial positions between sequencing spots and generate unified comprehensive representations through Graph Neural Networks.
For instance, the spatial transcriptomic method STAGATE~\cite{dong2022deciphering} utilizes KNN to construct a spatial adjacency graph for encoded transcriptomics data.
Spatial multi-modal omics method SpatialGlue~\cite{long2024deciphering} constructs the omics KNN graph as well as spatial adjacency graph for each modality separately and obtains a comprehensive latent representation through GCN~\cite{kipf2016semi}.
% Recently, graph-based encoders that encode omics features through correlation graph structure have become the consensus choice for spatial multi-modal omics methods due to their excellent performance~\cite{dong2022deciphering,hu2021spagcn,long2023spatially}.
However, these methods ignore the interference in semantic relations caused by the perturbations inevitably introduced during the biological sequencing process, which KNN fails to overcome due to the artificially set K value limit.

\begin{figure}
  \centering
  \includegraphics[width=0.45\textwidth]{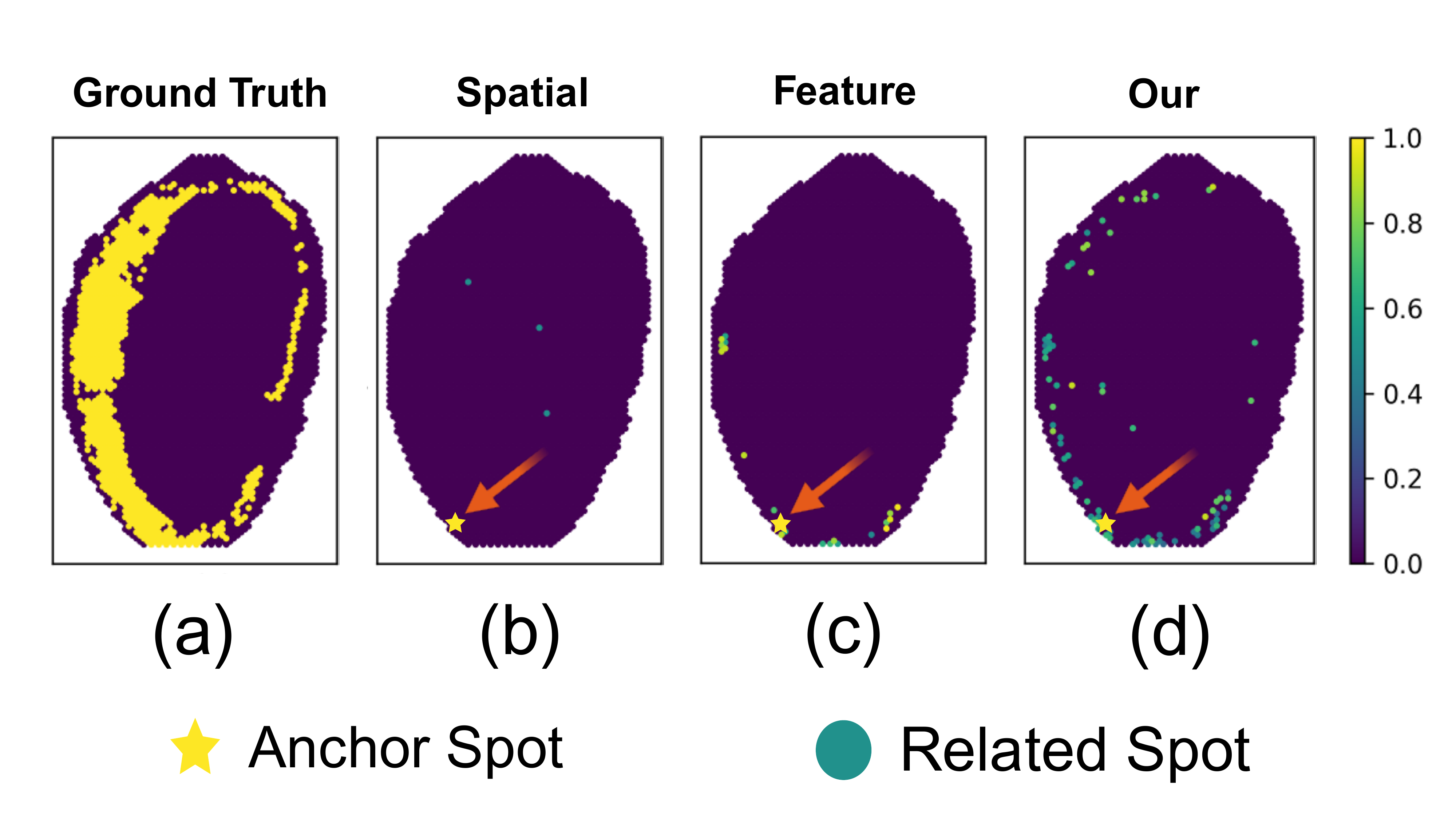}
  \caption{Visualization of adjacency graph of random anchor spots on the Human Lymph Node dataset. (a) highlights the spots with the same cell type as the anchor spots. (b) and (c) shows the related spots found by spatial and feature adjacency K-nearest neighbor (KNN) graphs, respectively. (d) shows the potential relevant spots revealed by our proposed dynamic omic-specific graph.
  }
  \label{intro}
\end{figure}

Technically, the current single-cell sequencing technology inevitably introduces perturbations in the sequencing data due to biological variation and other uncontrollable factors.
These perturbations hide some informative semantic relations, resulting in incomplete modeling of semantic relations by the fixed KNN graph.
To verify this intuition, we follow SpatialGlue~\cite{long2024deciphering} to perform KNN to construct the RNA adjacency graph in the Human Lymph Node dataset and randomly select a sequencing spot as an anchor for visualization.
As shown in Figure~\ref{intro}, compared with Ground Truth (GT), the spatial (Figure~\ref{intro} (b)) and feature (Figure~\ref{intro} (c)) adjacency graphs constructed by KNN only reveal limited connections and are limited by the spatial region obviously, \textit{e.g.}, the upper right area has no relevant feature points recognized in Figure~\ref{intro} (c), although spots with the same type to the anchor exist there.
This fixed graph structure with severe semantic information loss inevitably causes undesired performance degradation of spatial multi-omics resolved models.
To address this problem, we propose an omics-specific dynamic graph architecture to denoise perturbations by learning cross-modal knowledge and thereby reveal latent semantic relations.
Compared with the graph constructed by KNN, our method can learn a more informative graph structure (Figure~\ref{intro} (d)) that is closer to GT.

In addition, sequencing spot annotations and the number of spot types are usually unknown in practical scenarios, which makes it difficult to provide type-related knowledge for the optimization of omics-specific graphs.
Inspired by Bayesian Mixture Models~\cite{ronen2022deepdpm,chang2013parallel,zhao2023learning}, we propose a dynamic prototype contrastive learning method to address this issue.
Benefiting from the adaptability of the Bayesian Gaussian Mixture Model to the number of clusters in an open Bayesian environment, the dynamic prototype contrast learning can adaptively perceive the number of cell types and optimize the learnable graph architecture to reveal potential correlations among spots.

In this paper, we propose a novel spatial multi-omics resolved framework, termed \textbf{PR}ototype-\textbf{A}ware \textbf{G}raph \textbf{A}daptive Aggregation for Spatial Multi-modal Omics Analysis (PRAGA).
PRAGA learns an adaptive omics-specific graph to model spatial neighborhoods as well as latent semantic relations among spots.
Moreover, a dynamic prototype contrastive learning method is proposed to optimize omics-specific graphs despite the unknown number of spot types.
Extensive qualitative and quantitative experimental results demonstrate that PRAGA significantly outperforms state-of-the-art methods in aggregating spatial multi-modal omics information into spot-type-resolvable representations.
Our contributions are summarised as follows:
\begin{itemize}
\item We propose a novel spatial multi-modal omics resolved framework PRAGA for aggregating multi-modal omics data with their corresponding spatial positions.
\item We focus on latent semantic relations hidden by sequencing perturbations, which KNN fails to capture, and propose dynamic omics-specific graphs to learn these semantic relations from other omics modalities.
\item We propose a learnable spatial aggregation graph structure to adaptively aggregate features and spatial information to obtain omics-specific encodings.
\item We consider the common lack of biological priors in practical scenarios and propose a dynamic prototype contrastive learning to optimize PRAGA by adaptively sensing the number of sequencing point types.
% \item We conduct qualitative and quantitative validation experiments across four real datasets and one simulated dataset, demonstrating that our proposed PRAGA achieves State-Of-The-Art performance.

\end{itemize}

\section{Related work}
\subsection{Multi-omics Aggregation}
Multi-modal omics aggregation aims to integrate multiple omics data from the same biological sample to analyze gene expression and regulatory processes comprehensively~\cite{wu2022megps,zheng2024sccat}.
Existing multi-modal omics aggregation methods can be mainly divided into three categories: 1) non-negative matrix factorization; 2) Bayesian statistics; and 3) deep learning methods.
The non-negative matrix factorization decomposes multi-modal omics data into a common factor matrix to obtain a unified representation~\cite{kim2020citefuse}.
% However, expansion of the data scale will lead to an exponential increase in the computational burden of matrix decomposition, which makes it unsuitable for large-scale omics data aggregation.
Bayesian statistics methods fit one omics data to the conditional probability distribution of other omics but require biological priors~\cite{argelaguet2020mofa+}.
Compared with the above two methods, deep learning methods have attracted widespread attention due to their scalability and excellent performance.
TotalVI~\cite{gayoso2021joint} jointly models RNA and protein data into a low-dimensional space based on a variational autoencoder to obtain the comprehensive representation.
MultiVI~\cite{ashuach2023multivi} performs independent encoders for each modality of omics data and maps the multi-modal encodings to a joint latent space.
Despite integrating multi-modal omics, these methods ignore the impact of spatial location on omics features.

\subsection{Spatial Resolved Omics}
Recently, spatial resolved omics technologies, represented by spatial transcriptomics~\cite{moses2022museum}, have been proposed to associate spatial information with transcriptomic data.
STAGATE~\cite{dong2022deciphering} captures the local structure and spatial dependency of transcriptomic data via a graph attention network~\cite{velivckovic2017graph}.
PAST~\cite{li2023latent} performs a transformer architecture to capture self-similarity and global dependencies in transcriptomic data.
The most advanced work SpatialGlue~\cite{long2024deciphering} constructs a K-nearest neighbor (KNN) graph to model semantic relations and integrate spatial information with multi-modal omics through attention weights to provide a comprehensive representation.
However, due to the inherent perturbations introduced during sequencing, the semantics of some omics features are disturbed and difficult to model by KNN graphs.
In this paper, we propose a dynamic graph to mitigate the interference of perturbations and capture latent semantic relations through cross-modal knowledge and dynamic prototype contrastive learning.

\section{Method}

\begin{figure*}
  \centering
  \includegraphics[ width=0.90\textwidth]{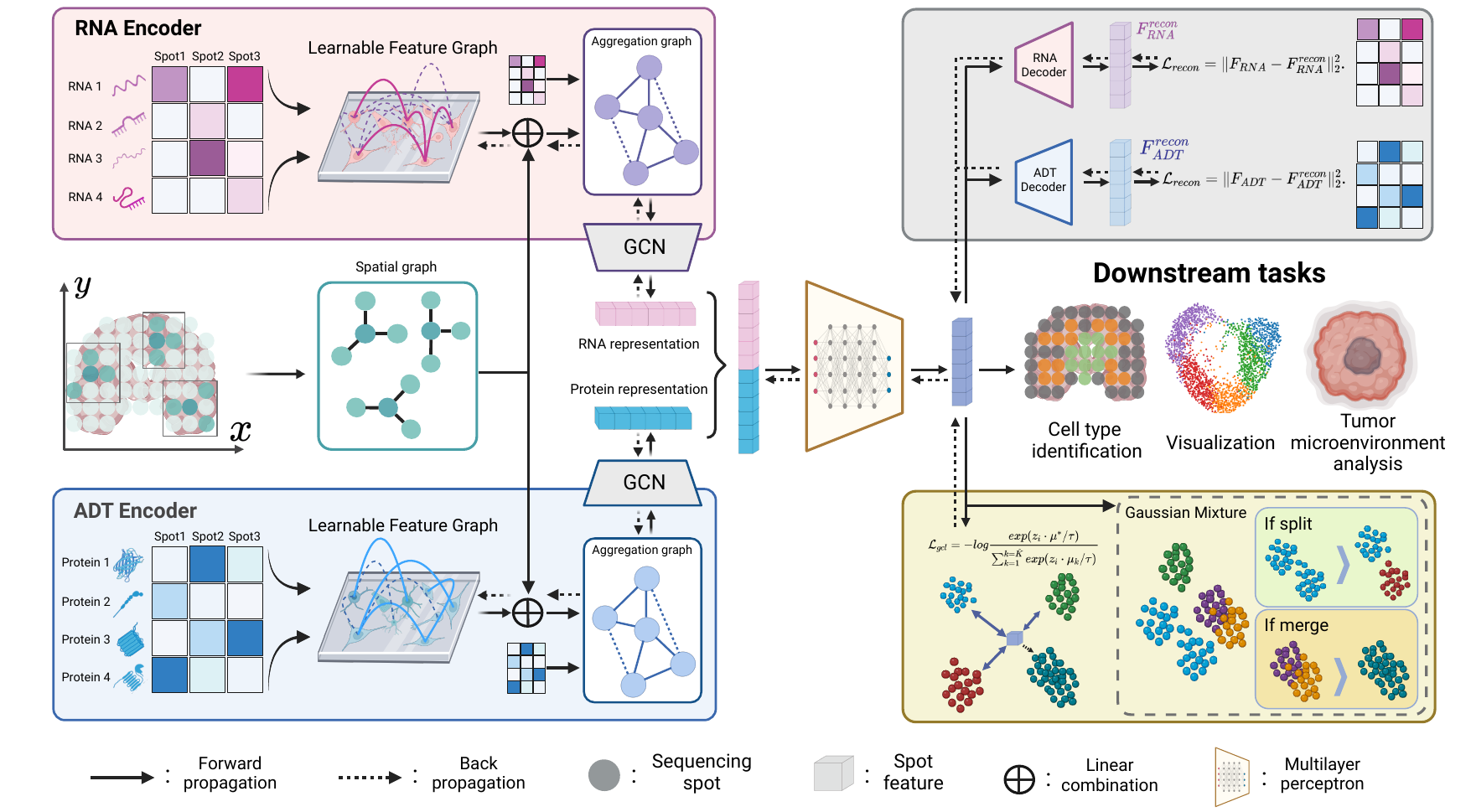}
  \caption{The framework of the proposed PRAGA. 
  A learnable feature graph is used to explore the potential correlations between spots.
  The comprehensive representations of multi-modality (RNA, Protein) are obtained by aggregating modality-specific encodings, which GCNs calculate with linear combinations of learnable feature graphs and spatial adjacency graphs.
  The entire model is trained with a modality-specific reconstruction loss and a dynamic cluster prototype contrastive loss for the latent representation, where clusters are obtained via a Gaussian Mixture Model and optimized via split and merge operations.
  }
  \label{_frame}
\end{figure*}

% The potential proximity relationships of multimodal features (RNA, ADT) are represented by learnable feature graphs. The spatial adjacency graph is constructed through spatial coordinate neighbors. 

% In this section, we first formulate spatial multi-modal omics resolve tasks.
% Then, we propose a dynamic omic-specific graph structure to learn potential semantic relations between sequencing spots.
% Further, we design a spatial multi-modal omics autoencoder with dynamic omic-specific graphs to aggregate spatial and multi-modal omics information.
% Finally, we propose a dynamic prototype contrastive learning to optimize learnable feature graphs in the setting with an unknown number of cell types.
% An overview of our proposed PRAGA is illustrated in Figure~\ref{_frame}.
% The overall process of our proposed PRAGA is summarized in Algorithm~\ref{alg}. 

\subsection{Preliminaries}

Spatial multi-modal omics resolve tasks aim to integrate spatial information and multi-modal omics data, such as RNA sequencing in transcriptomics, Assay for Transposase-Accessible Chromatin (ATAC) in genomics, and Antibody-Derived Tags (ADT) in proteomics, to obtain a unified latent representation.
Given spatial coordinates of $N$ sequencing spots $\boldsymbol{S}=\left\{(x_i, y_i)\right\}_{i=1}^{N}$ and their corresponding features from $M$ modalities $\boldsymbol{F_M}=\left\{\boldsymbol{f_{i}^{m}}\right\}_{i=1,m=1}^{i=N,m=M}$, where $\boldsymbol{f^{m}}\in\mathbb{R}^{D_m}$ represent the $D_m$-dimensional features of the $m$ modality, usually obtained by data preprocessing such as principal component analysis~\cite{pearson1901liii} and highly variable gene screening.
Spatial multi-modal omics resolve methods provide a function $\Phi$ to aggregate $\boldsymbol{F_M}$ and $\boldsymbol{S}$ to $D_z$-dimensional comprehensive latent representation $\boldsymbol{\mathcal{Z}}$:
\begin{equation}\label{phi}
    \boldsymbol{\mathcal{Z}} = \Phi\left(\boldsymbol{F_{M}}, \boldsymbol{S}\right).
\end{equation}

The comprehensive representation $\boldsymbol{\mathcal{Z}}$ can be applied to downstream bioinformatics analysis tasks such as cell identification~\cite{littman2021joint}, tumor microenvironment analysis~\cite{hunter2021spatially,janesick2023high}, $\textit{etc}$.

\subsection{PRAGA}
In this paper, we propose a parameterized architecture PRAGA as the function $\Phi$ in Eq.~(\ref{phi}) to integrate multi-modal omics data and spatial information into a comprehensive latent representation.
Specifically, we first construct a dynamic omic-specific feature graph $\mathcal{G}^{F}_m=(\mathcal{V}^{F}_m, \mathcal{E}^{F}_m)$ for each modality, where $\mathcal{V}$ is the set of graph nodes and $\mathcal{E}$ is the set of undirected edges.
For the convenience of formulation, we express $\mathcal{V}_m$ and $\mathcal{E}_m$ in matrix form as the omic-specific feature matrix $\boldsymbol{F_m}$ and adjacency matrix $\boldsymbol{A^F_m}$, $\textit{i.e.}$, $\mathcal{G}^{F}_m=(\boldsymbol{F_m}, \boldsymbol{A^F_m})$.
Then, we construct a K-Nearest Neighbor (KNN) spatial adjacency graph $\mathcal{G}^{S}=(\boldsymbol{S}, \boldsymbol{A^S})$ and learn an aggregation graph $\hat{\mathcal{G}}^{F}_m=(\boldsymbol{F_m}, \boldsymbol{\hat{A}^F_m})$ by combining the spatial graph $\mathcal{G}^{S}$ and the feature graph $\mathcal{G}^{F}_m$.
Further, the GCN is performed as an encoder to encode omic features with the spatial aggregation graph $\hat{\mathcal{G}}^{F}_m$ for each modality.
Finally, the unified comprehensive representations $\boldsymbol{\mathcal{Z}}$ are integrated from all modalities by a Multi-Layer Perceptron.

To mitigate the impact of perturbations on a single modality, we propose a reconstruction loss, $\mathcal{L}_{recon}$, using a modality-specific decoder to impart cross-modal knowledge to dynamic omic-specific graphs. Facing the challenge of unknown biological priors in practice, we propose dynamic prototype contrastive learning, which adaptively determines the number of spot types and uncovers latent semantic relationships between sequencing spots. An overview of our proposed PRAGA method is illustrated in Figure~\ref{_frame}.
The overall process of our proposed PRAGA is summarized in Algorithm~\ref{alg}.

\subsection{Dynamic Omic-specific Graph}
Limited by uncontrollable factors such as biological variation, some single-cell sequencing data are inevitably disturbed by perturbations.
In this case, performing KNN to model the inter-spot semantic relations is simplicity and crudity, since it discards some informative semantic relations, especially the latent relations disturbed by sequencing perturbations.
Here we provide a more fine-grained solution: building a dynamic graph structure to learn latent semantic relations while resisting sequencing perturbations by learning cross-modal knowledge.

% State-of-the-art methods construct K-Nearest Neighbor (KNN) graphs to model the semantic relations between sequencing spots and generate omics-specific representations through Graph Neural Networks.
% However, the KNN graph, limited by the artificially selected K value, exhibits simplicity and crudity, since it inevitably discards some informative semantic relations, especially the potential relations disturbed by sequencing perturbations.
% Here we provide a more fine-grained solution: building a learnable dynamic graph structure to adaptively learn latent semantic relations while resisting the impact of sequencing perturbations.

Given a specific omics modality, such as RNA sequencing data $\boldsymbol{F_{\text{RNA}}}$ with $N$ sequencing spots, we construct a learnable parameter matrix $\boldsymbol{A^F_{\text{RNA}}}\in\mathbb{R}^{N\times N}$ to model the inter-spots semantic relation graph $\mathcal{G}^F_{\text{RNA}}=(\boldsymbol{F_{\text{RNA}}}, \boldsymbol{A^F_{\text{RNA}}})$.
The KNN undirected graph is used to initialize $\mathcal{G}^F_{\text{RNA}}$ to ensure initial sparsity.
Specifically, for sequencing spot $i$, we set K spots with the closest Euclidean distance to its sequencing feature $\boldsymbol{f^{\text{RNA}}_{i}}$ as neighbors, where K is set to 20 following~\cite{long2024deciphering}.
Then we initialize $A^F_{\text{RNA},i,j}$=1 if and only if sequencing spot $j$ is a neighbor of $i$, otherwise $A^F_{\text{RNA},i,j}$=0.
The intuition behind the design of this KNN-initialized dynamic graph is to adjust the KNN graph by learning new edge weights to model the latent semantic relations disturbed by sequencing perturbations.

\subsection{Spatial Aggregation Encoding}\label{Sec3.3}
Equipped with the dynamic omic-specific graph, we construct a spatial aggregation graph to encode omics data by combining feature and spatial information.
We first initialize the learnable feature graph $\mathcal{G}^F_{m}$ and spatial adjacency graph $\mathcal{G}^S=(\boldsymbol{F_{m}}, \boldsymbol{A^S})$ through KNN, where $\boldsymbol{A^S}\in\mathbb{R}^{N\times N}$ is the spatial adjacency matrix calculated based on spatial coordinates, the subscript $m$ present specific modality and $m\in\{\text{RNA, ADT, ATAC}\}$ in our setting.
The spatial aggregated graph $\hat{\mathcal{G}}_{m}=(\boldsymbol{F_{m}}, \boldsymbol{\hat{A}_{m}})$ is obtained by combining $\mathcal{G}^S$ and $\mathcal{G}^F_{m}$ with learnable parameters:

\begin{equation}\label{hatA}
    \boldsymbol{\hat{A}_{m}}= w^S_{m} \boldsymbol{A^S} + w^F_{m} \boldsymbol{A^F_{m}},
\end{equation}
where $w^S_{m}$ and $w^F_{m}$ are learnable parameters. 
Then, we perform a one-layer GCN~\cite{kipf2016semi} as an encoder to encode sequencing features $\boldsymbol{F_{m}}$ with spatial aggregated graph $\hat{\mathcal{G}}_{m}$:
\begin{equation}\label{Z}
    \boldsymbol{Z_{m}} = \text{GCN}^{en}_{m}(\boldsymbol{F_{m}},\boldsymbol{\hat{A}_{m}})=\boldsymbol{\hat{A}_{m}} \boldsymbol{F_{m}} \boldsymbol{W_{m}^{en}},
\end{equation}
where $\boldsymbol{W_{m}^{en}}$ is a learnable parameter matrix in the encoder.

After obtaining modality-specific encoding for all modalities $\{\boldsymbol{Z_{\text{RNA}}}, \boldsymbol{Z_{\text{ADT}}}, \boldsymbol{Z_{\text{ATAC}}}\}$, a Multi-Layer Perceptron (MLP) is utilized to map differential modality encoding to a unified comprehensive representation:
\begin{equation}\label{MLP}
    \boldsymbol{\mathcal{Z}} = \text{MLP}(\text{Concat}(\{\boldsymbol{Z_{\text{RNA}}}, \boldsymbol{Z_{\text{ADT}}}, \boldsymbol{Z_{\text{ATAC}}}\})),
\end{equation}
where $\text{Concat}(\cdot)$ is the concatenation operation.

% Modality-specific decoders are performed to constrain the retention of modality-specific information during aggregation.

For each modality, we perform a modality-specific decoder, also implemented by a single-layer GCN, to reconstruct omics features from comprehensive encoding $\boldsymbol{\mathcal{Z}}$ with spatial adjacency graph $\mathcal{G}^S$:
\begin{equation}
    \boldsymbol{F_m^{recon}} = \text{GCN}^{de}_m(\boldsymbol{\mathcal{Z}},\boldsymbol{A^S})=\boldsymbol{A^S} \boldsymbol{\mathcal{Z}} \boldsymbol{W_m^{de}},
\end{equation}

The reconstructed loss can be calculated by the mean square error between reconstructed omics features and origin omics features:
\begin{equation}\label{Lrecon}
    \mathcal{L}_{recon}=\frac{1}{M}\sum_{m=1}^{M}w_m\| \boldsymbol{F_m} - \boldsymbol{F_m^{recon}} \|_2^2 .
\end{equation}
% Spatial multi-modal omics autoencoder encodes spatial information and features of each modality into a comprehensive representation $Z$. 
where $w_m$ represents the reconstruction weight of the $m$ modality.
The reconstruction loss $\mathcal{L}_{recon}$ constrains the encoding model to retain modality-specific information while providing cross-modal supervision to each omic-specific graph structure, so that the omic-specific graph can obtain knowledge from other modalities to mitigate the impact of sequencing perturbations and promote dynamic omic-specific graph to discover potential semantic relations.

Empirically, drastic changes in the adjacency graph present a risk of unstable training~\cite{jin2020graph}.
Therefore, we calculate homogeneity loss utilizing the F-norm of the interpolated feature graph before and after the update to constrain the change of the feature graph:
% \begin{equation}
% \mathcal{L}_{f} = \sqrt{\sum\nolimits_{i=1}^{N \times N} \left|\boldsymbol{A^{F}_{\text{RNA},i}} - \boldsymbol{A^{ref}_{\text{RNA},i}}\right|^2},
% \end{equation}

\begin{equation}\label{Lf}
\mathcal{L}_{h} = \frac{1}{M}\sum_{m=1}^{M}\|\boldsymbol{A^{F}_{m}} - \boldsymbol{A^{ref}_{m}}\|_{\mathcal{F}},
\end{equation}
where $\boldsymbol{A^{ref}_{m}}$ is a reference graph of modal $m$, initialized in the same way as $\boldsymbol{A^{F}_{m}}$, and slowly updated by exponential moving average during the learning process of $\boldsymbol{A^{F}_{m}}$:

\begin{equation}\label{EMA}
    \boldsymbol{A^{ref}_{m,e}} =
\begin{cases} 
\text{KNN }(\boldsymbol{F_{m}}), & e=1 \\
\alpha \boldsymbol{A^{ref}_{m,e-1}} + (1-\alpha)\boldsymbol{A^{F}_{m,e-1}}, & e>1

\end{cases}
\end{equation}
where $e$ is the training epoch index, and $\alpha$ is a hyper-parameter to control the moving speed.
Homogeneity loss constrains the omics-specific graph to learn only a small number of new associated edges each time, providing interpretability for the omics-specific graph while maintaining training stability.

\subsection{Dynamic Prototype Contrastive Learning}\label{Sec3.4}

In practice, spot type annotations and even the number of spot types are often unknown, hindering the existing methods from exploring inter-spot latent relations from the spot clustering perspective.
Fortunately, we draw inspiration from Bayesian Mixture Models~\cite{ronen2022deepdpm,chang2013parallel} and propose a Bayesian Gaussian Mixture Model-based dynamic prototype contrastive learning.

 For comprehensive representation $\boldsymbol{\mathcal{Z}}$ obtained by Eq.~(\ref{MLP}), we set an initial number of clusters $C$ and utilize the Gaussian Mixture Model to assign $\boldsymbol{\mathcal{Z}}$ into $C$ clusters.
Note that the parameter sensitivity experiments verify the insensitivity of our method to initial $C$ values, so only a rough $C$ value is needed in practical scenarios.
 The mean and the number of examples in each cluster are formulated as $\boldsymbol{\mu_c}$ and $N_c$, where the subscript $c$ represents the cluster index.
 To dynamically adjust the number of clusters, each cluster is further divided into two sub-cluster, whose mean and the number of examples are denoted as $\boldsymbol{\mu_{c,s}}$ and $N_{c,s}$, where $s\in\{1,2\}$ is the sub-cluster index.
Then, we set a split criterion for each cluster to decide whether this cluster needs to split:

\begin{equation}
    \mathcal{S} = \frac{\Gamma(N_{c,1})L(\boldsymbol{Z_{c,1}}, \nu, \kappa)\Gamma(N_{c,2})L(\boldsymbol{Z_{c,2}}, \nu, \kappa)}{\Gamma(N_{c})L(\boldsymbol{Z_{c}}, \nu, \kappa)},
\end{equation}
where $\Gamma(\cdot)$ is the Gamma function, $L(\cdot, \nu, \kappa)$ is the marginal likelihood with a Normal Inverse Wishart (NIW) distribution as the prior, $ \nu$ and $ \kappa$ are hyper-parameters of NIW. 
If $\mathcal{S}>1$, the original cluster will be replaced by one of its sub-clusters, and the other sub-cluster is added as a new cluster:

\begin{equation}
   \boldsymbol{\mu_c} := \boldsymbol{\mu_{c,1}},
    \boldsymbol{\mu_{\textit{C+}1}} := \boldsymbol{\mu_{c,2}}.
\end{equation}

Similarly, a merge criterion is set to determine whether two clusters $i$ and $j$ need to be merged into one:
\begin{equation}
   \mathcal{M} = \frac{\Gamma(N_{i}+N_{j})L(\boldsymbol{Z_{i}}\cup \boldsymbol{Z_{j}}, \nu, \kappa)}{\Gamma(N_{i})L(\boldsymbol{Z_{i}}, \nu, \kappa)\Gamma(N_{j})L(\boldsymbol{Z_{j}}, \nu, \kappa)}.
\end{equation}
The new merge cluster will replace the original two clusters with the average of the two clusters if $\mathcal{M}>1$:
\begin{equation}
    \boldsymbol{\mu_{i}}:=\boldsymbol{\varnothing},\boldsymbol{\mu_{j}}:=\boldsymbol{\varnothing},\boldsymbol{\mu_{\textit{C}\text{-}2\text{+}1}} := \frac{\boldsymbol{\mu_{i}} + \boldsymbol{\mu_{j}}}{2}
\end{equation}

% Thanks to the adaptive optimization of the number of clusters in split and merge operations, our method can overcome the dilemma of an unknown number of cell types in practice and explore the potential relations .
After obtaining the adjusted clusters, we assign the nearest cluster to each sequencing spot, and perform contrastive learning to further optimize learnable feature graphs with cluster centers as prototypes:

\begin{equation}\label{Ldpcl}
    \mathcal{L}_{dpcl}=-\frac{1}{N}\sum_{\boldsymbol{z_i} \in \boldsymbol{\mathcal{Z}}}{log\frac{exp\left(\boldsymbol{z_i} \cdot \boldsymbol{\mu^{\ast}}/ \tau \right)}{\sum_{c=1}^{c=\hat{C}}{exp\left(\boldsymbol{z_i} \cdot \boldsymbol{\mu_c}/ \tau \right)}}},
\end{equation}
where  $\boldsymbol{\mu^{\ast}}$ is the center of the cluster closest to $\boldsymbol{z_i}$, $\hat{C}$ is the number of clusters updated after the spilt and merge operations, $\tau$ is the temperature hyper-parameter.
Thanks to the split and merge operations, the $\mathcal{L}_{dpcl}$ can obtain an adaptive number of prototypes for contrastive learning, enabling the model to generate a resolvable comprehensive representation even if the number of categories of spots is unknown.
In addition, supervision from $\mathcal{L}_{dpcl}$ also drives omic-specific graphs to learn latent semantic relations, thereby obtaining a informative omic-specific encoding.

The total loss of our method is the combination of the homogeneity loss, reconstructed loss, and contrastive learning:

\begin{equation}\label{Ltotal}
    \mathcal{L}_{Total}=\mathcal{L}_{h} + \mathcal{L}_{recon} + \beta\mathcal{L}_{dpcl},
\end{equation}
where $\beta$ is the weight hyper-parameter that balances the loss component $\mathcal{L}_{dpcl}$.

Through the joint constraints of three losses $\mathcal{L}_{f}$, $\mathcal{L}_{recon}$, and $\mathcal{L}_{dpcl}$, our proposed dynamic omics-specific graph is able to resist the interference of sequencing perturbations by learning semantic knowledge from other modalities and modeling abundant semantic relations.
On this basis, our proposed PRAGA framework integrates spatial information and multi-modal omics features into a unified comprehensive representation in an end-to-end manner.

\begin{algorithm}[tb]
    \caption{PRAGA}
    \label{alg}
    \textbf{Input}: Multi-modal omics data, \textit{e.g.}, RNA sequencing $F_{\text{RNA}}$, Chromatin accessibility $F_{\text{ATAC}}$, Protein expression  $F_{\text{ADT}}$, and their shared spatial coordinates $S$\\
    \textbf{Parameter}: The total epoch $E$, a initial K, temperature coefficient $\tau$, and loss weight $\beta$.
    \begin{algorithmic}[1]
        \STATE Initialize the dynamic graph $\mathcal{G}^{F}=(\boldsymbol{F}, \boldsymbol{A^F})$ for each modality and spatial position using K-Nearest Neighbor. \\
        % $A_{\text{RNA}}=KNN(F_{\text{RNA}})$, $A_{\text{ATAC}}=KNN(F_{\text{ATAC}})$,\\
        % $A_{\text{ADT}}=KNN(F_{\text{ADT}})$,
        % $A_{\text{S}}=KNN(S)$,
        \FOR{$e$ in 1,2,...,$E$}
            \STATE Update reference graph for each modality in Eq.~(\ref{EMA}).
            \STATE Calculate the $\mathcal{L}_h$ for each modality in Eq.~(\ref{Lf}).
            \STATE Construct the aggregated graph $\hat{\mathcal{G}}^{F}=(\boldsymbol{F}, \boldsymbol{\hat{A}^F})$ for each modality in Eq.~(\ref{hatA}).
            \STATE Obtain modality-specific encodes $\boldsymbol{Z}_{\text{RNA}}$, $\boldsymbol{Z}_{\text{ATAC}}$, $\boldsymbol{Z}_{\text{ADT}}$ in Eq.~(\ref{Z}).
            \STATE Mapping all modality-specific encodes to a unified comprehensive representation $\boldsymbol{\mathcal{Z}}$ in Eq.~(\ref{MLP}).
            \STATE Calculate reconstruction loss $\mathcal{L}_{recon}$ in Eq.~(\ref{Lrecon}).
            \STATE Calculate dynamic prototype contrastive learning loss $\mathcal{L}_{dpcl}$ in Eq.~(\ref{Ldpcl}).
            \STATE Train PRAGA by minimizing $\mathcal{L}_{Total}$ in Eq.~(\ref{Ltotal}).
        \ENDFOR
    \end{algorithmic}
    \textbf{Output} Unified comprehensive code $\boldsymbol{\mathcal{Z}}$
\end{algorithm}

\section{Experiments}

\begin{figure}[ht]
  \centering
  \includegraphics[width=0.48\textwidth]{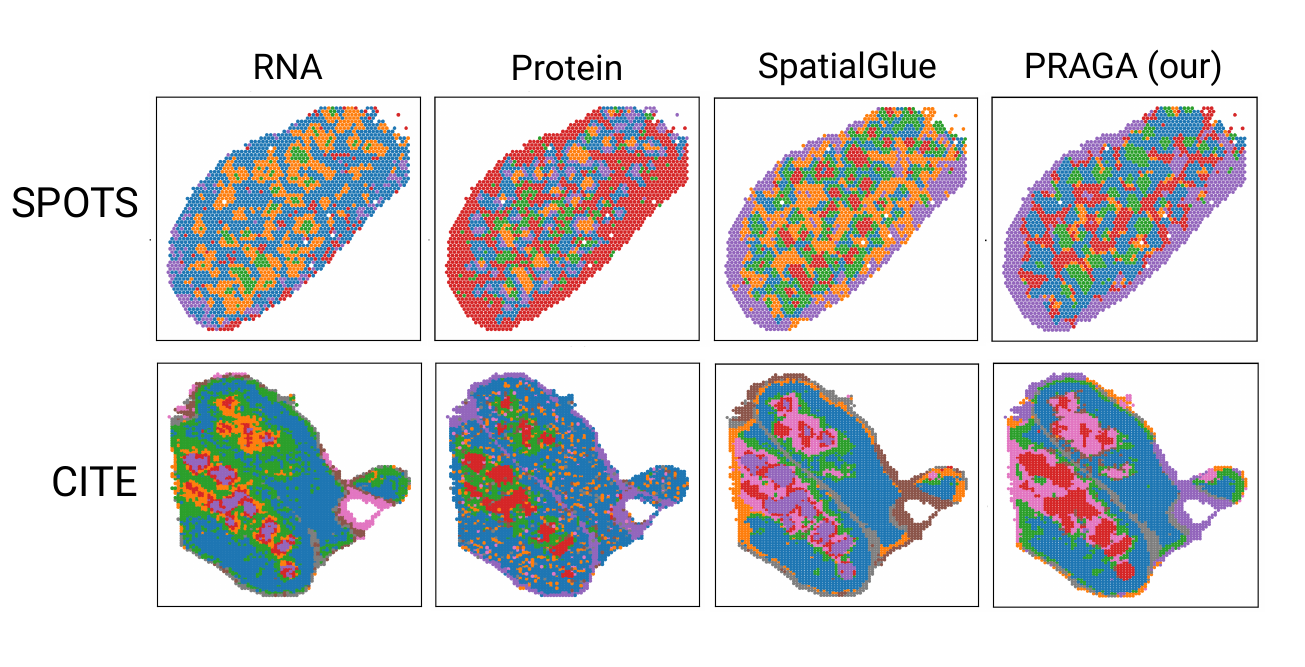}
  \caption{Visualization of qualitative experimental results on real datasets SPOTS mouse spleen (SPOTS) and mouse thymus stereo-CITE-seq (CITE).}
  \label{vis}
\end{figure}

\begin{table*}[tb]
\fontsize{9}{10}\selectfont
\centering
\setlength{\tabcolsep}{2.4mm}
\begin{tabular}{lccccccccc}
\toprule
Methods & MI(\%) & NMI(\%) & AMI(\%)  & FMI(\%) & ARI(\%) & V-Measure(\%) & F1-Score(\%) & Jaccard(\%)  & Compl.(\%)\\
\midrule
\multicolumn{10}{c}{Human Lymph Node Dataset}\\
\midrule
MOFA+ & 65.06 & 34.91 & 34.49 & 37.67 & 22.73 & 34.91 & 36.96 & 22.67 & 31.88\\
CiteFuse & 45.35 & 23.34 & 22.86 & 27.57 & 12.51 & 23.38 & 26.15 & 15.04 & 26.88\\
TotalVI & 25.51 & 14.72 & 14.14 & 26.59 & 6.45 & 14.72 & 26.55 & 15.31 & 15.15\\
MultiVI & 12.03 & 7.01 & 6.43 & 26.16 & 3.73 & 7.01 & 26.15 & 15.04 & 7.13\\
STAGATE & 1.42 & 0.79 & 0.12 & 20.83 & 0.22 & 0.79 & 20.73 & 11.56 & 0.74\\
PAST & 58.82 & 33.60 & 33.14 & \underline{41.42} & \underline{24.64} & 33.60 & \underline{41.41} & \underline{26.11} & 32.42\\
SpatialGlue & \underline{66.52} & \underline{36.07} & \underline{35.65} & 39.16 & 23.83 & \underline{36.07} & 38.79 & 24.06 & \underline{33.24}\\
PRAGA & \textbf{73.00} & \textbf{39.47} & \textbf{39.07} & \textbf{42.69} & \textbf{28.28} & \textbf{39.47} & \textbf{42.33} & \textbf{26.76} & \textbf{36.29}\\
$\Delta$ & +6.48 & +3.40 & +3.42 & +1.27 & +3.64 & +3.40 & +0.92 & +0.65 & +3.05\\
\midrule
\multicolumn{10}{c}{Mouse Brain Dataset}\\
\midrule
MOFA+ & 19.58 & 8.64 & 8.38 & 15.59 & 4.39 & 8.64 & 15.59 & 8.45 & 8.91\\
CiteFuse & 47.48 & 19.46 & 19.05 & 17.96 & 8.24 & 19.46 & 17.89 & 9.82 & 18.63\\
MultiVI & 17.88 & 8.47 & 8.22 & 18.12 & 3.81 & 8.47 & 17.58  & 9.63 & 9.46 \\
STAGATE & 48.45 & 21.25 & 21.03 & 22.36 & 12.21 & 21.25 & 22.36 & 12.59 & 21.76\\
PAST & 69.49 & 29.13 & 28.76 & 24.54 & 14.63 & 29.13 & 24.54 & 13.99 & 28.50\\
SpatialGlue & \underline{95.54} & \underline{37.83} & \underline{37.53} & \underline{33.78} & \underline{26.33} & \underline{37.83} & \underline{33.01} & \underline{19.77} & \underline{35.14}\\
PRAGA & \textbf{95.55} & \textbf{39.37} & \textbf{39.06} & \textbf{35.07} & \textbf{27.06} & \textbf{39.37} & \textbf{35.02} & \textbf{21.23} & \textbf{37.88}\\
$\Delta$ & +0.01 & +1.54 & +1.53 & +1.29 & +0.73 & +1.54 & +2.01 & +1.46 & +2.74\\
\midrule
\multicolumn{10}{c}{Spatial Multi-modal Omics Simulation Dataset}\\
\midrule
MOFA+ & 1.02 & 0.58 & -0.23 & 21.32 & 0.39 & 0.58 & 21.27 & 11.90 & 0.52\\
CiteFuse & 1.23 & 0.66 & -0.10 & 17.17 & 0.03 & 0.66 & 16.56 & 9.03 & 0.57\\
TotalVI & 1.36 & 0.72 & -0.02 & 15.93 & -0.09 & 0.72 & 15.03 & 8.12 & 0.61\\
MultiVI & 1.22 & 0.77 & -0.05 & 25.20 & -0.01 & 0.77 & 25.05 & 14.32 & 0.75\\
STAGATE & 7.40 & 3.91 & 3.91 & 17.25 & 1.56 & 3.91 & 16.23 & 8.83 & 3.31\\
PAST & 2.09 & 1.18 & -0.18 & 19.17 & 0.07 & 1.18 & 18.91 & 10.44 & 1.05\\
SpatialGlue & \underline{150.13} & \underline{96.98} & \underline{96.97} & \underline{98.21} & \underline{97.69} & \underline{96.98} & \underline{98.21} & \underline{96.48} & \underline{96.95}\\
PRAGA & \textbf{153.35} & \textbf{99.06} & \textbf{99.05} & \textbf{99.47} & \textbf{99.32} & \textbf{99.06} & \textbf{99.47} & \textbf{98.95} & \textbf{99.08}\\
$\Delta$ & +3.22 & +2.08 & +2.08 & +1.26 & +1.63 & +2.08 & +1.26 & +2.47 & +2.13\\
\bottomrule
\end{tabular}

\caption{Quantitative experimental results with nine metrics for the human lymph node dataset, the spatial epigenome–transcriptome mouse brain dataset, and the spatial multi-omics simulation dataset. The symbol $\Delta$ indicates the performance improvement of our proposed PRAGA over the best comparison method. Underline identifies the second-best results. The best experimental results are marked in bold.}

\label{table1}

\end{table*}

\subsection{Exprimental Setups}

\subsubsection{Datasets.}
We conduct quantitative and qualitative experiments on five public datasets to verify the effectiveness of the proposed method: 1) Human Lymph Node dataset~\cite{long2024deciphering}; 2) Spatial epigenome–transcriptome mouse brain dataset~\cite{zhang2023spatial}; 3) Mouse thymus stereo-CITE-seq dataset~\cite{liao2023integrated}; 4) SPOTS mouse spleen dataset~\cite{ben2023integration}; 5) Spatial multi-modal omics simulation datasets~\cite{long2024deciphering}.
These datasets are detailed in the Appendix\footnote{https://arxiv.org/abs/2409.12728}.

\subsubsection{Baselines.}
We compare our approach with recent advanced works, include 4 multi-modal omics methods, MOFA+~\cite{argelaguet2020mofa+}, MultiVI~\cite{ashuach2023multivi}, TotalVI~\cite{gayoso2021joint}, CiteFuse~\cite{kim2020citefuse}, 2 spatial transcriptomic methods, STAGATE~\cite{dong2022deciphering}, PAST~\cite{li2023latent}, and the latest spatial multi-omics work SpatialGlue~\cite{long2024deciphering}.

% \subsubsection{Implementation Settings.}
% We train our PRAGA using a stochastic gradient descent optimizer with a learning rate of 0.01 and a momentum of 0.9.
% The total training epochs are set to 300 for the spatial epigenome–transcriptome mouse brain dataset and 200 for the other datasets.

% In terms of hyper-parameters, we set the exponential moving average speed $\alpha$ to 0.9 for stable training.
% The initial number of cell categories is set to a value close to reality, which is 18 for the spatial epigenome–transcriptome mouse brain dataset, 10 for the Human Lymph Node and Stereo-CITE-seq mouse thymus, and 5 for the simulated dataset and SPOT dataset.
% The results of parameter sensitivity experiments demonstrate that our method is robust to the selection of the above hyperparameters.

\subsubsection{Metrics}
We selected 9 different metrics to evaluate the performance of the model, including Mutual information (MI), Normalized Mutual Information (NMI), Adjusted Mutual Information (AMI), Fowlkes-Mallows Index (FMI), Adjusted Rand Index (ARI), Variation of Information Measure (V-Measure), F1-Score, Jaccard Similarity Coefficient (Jaccard), and Completeness (Compl.).
Detailed experimental settings are provided in the Appendix.
% The calculation process of these metrics is shown in the Appendix.

\begin{table*}[t]
\fontsize{9}{10}\selectfont
\centering
\setlength{\tabcolsep}{2.4mm}
\begin{tabular}{lccccccccc}
\toprule
Methods & MI(\%) & NMI(\%) & AMI(\%)  & FMI(\%) & ARI(\%) & V-Measure(\%) & F1-Score(\%) & Jaccard(\%)  & Compl.(\%)\\
\midrule
PRAGA & \textbf{73.00} & \textbf{39.47} & \textbf{39.07} & \textbf{42.69} & \textbf{28.28} & \textbf{39.47} & \textbf{42.23} & \textbf{26.76} & \textbf{36.29}\\
\midrule
KNN & 71.72 & 38.69 & 38.23 & 42.49 & 28.06 & 38.69 & 42.02 & 26.60 & 35.50\\
w/o $\mathcal{L}_{h}$ & 72.26 & 39.25 & 38.85 & 42.56 & 27.96 & 39.25 & 42.17 & 26.72 & 36.22\\
w/o $\mathcal{L}_{recon}$ & 64.44 & 35.15 & 34.73 & 38.85 & 23.83 & 35.15 & 38.32 & 23.7 & 32.14 \\
w/o $\mathcal{L}_{dpcl}$ & 70.96 & 38.46 & 38.05 & 42.35 & 27.81 & 38.46 & 41.91 & 26.51 & 35.51\\
\bottomrule
\end{tabular}

\caption{Ablation study for our proposed learnable omics-specific graph, homogeneity loss $\mathcal{L}_{h}$, reconstruction loss $\mathcal{L}_{recon}$, and dynamic prototype contrast learning loss $\mathcal{L}_{dpcl}$ on the human lymph node dataset.
KNN represents using the K-Nearest Neighbor graph to replace our proposed learnable omics feature map.
w/o is the abbreviation of without.
The best experimental results are marked in bold.}

\label{Ablation}

\end{table*}

\subsection{Qualitative experimental results}
We first conducted qualitative experiments on the SPOTS mouse spleen (SPOTS) and mouse thymus stereo-CITE-seq (CITE) dataset to verify the aggregation effect of PRAGA on multi-modal omics data with spatial positions.
We cluster RNA, protein, and integrated encoding obtained by the state-of-the-art method SpatialGlue and our proposed PRAGA separately, and visualize them according to spatial position in Figure~\ref{vis}.
The visualization results show that spots of the same category integrated by PRAGA depict tighter and more continuous connections globally compared with SpatialGlue.
% On the one hand, the visualization results show that PRAGA can aggregate RNA and protein information while preserving spatial consistency.
% On the other hand, Compared with SpatialGlue, spots of the same category aggregated by PRAGA show tighter and more continuous connections globally.
We attribute this advantage to the learnable omic-specific graph structure and dynamic prototype contrastive learning, which enables PRAGA to resist perturbations and model latent semantic relations.
More qualitative experimental results are shown in the Appendix.

\subsection{Quantitative experimental results}
We conduct quantitative experiments on the human lymph node dataset, the spatial epigenome–transcriptome mouse brain dataset, and the spatial multi-modal omics simulation dataset.
Table~\ref{table1} summarizes the quantitative experimental results of these three datasets.
Note that for the spatial epigenome–transcriptome mouse brain dataset, the evaluation metrics reported in Table~\ref{table1} are based on the Antibody-Derived Tag (ADT) cluster labels as Ground Truth.
Benefiting from the dynamic graph structure and dynamic prototype contrastive learning, our method outperforms the baseline method on nine metrics consistently.
% We noticed that existing multi-modal omics and spatial transcriptomics methods perform poorly when processing spatial information and tri-modal omics information simultaneously (\textit{e.g.}, spatial multi-modal omics simulation datasets), while the recent spatial multi-modal omics method SpatialGlue and our proposed PRAGA show great advantages, which demonstrate the potential of spatial multi-modal omics methods in integrating spatial multi-modal omics data.
% We note that in the simulation dataset, which contains three modalities, spatial multi-omics methods including SpatialGlue and our PRAGA show great advantages over multi-omics methods and spatial transcriptomics methods, which demonstrate the potential of spatial multi-modal omics methods in integrating spatial multi-modal omics data.
Compared with the state-of-the-art work SpatialGlue, our proposed PRAGA achieves a significant performance improvement in F1-Score and NMI of $3.54\%$ and $3.40\%$ for the Human Lymph Node dataset, $2.01\%$ and $1.54\%$ for the Spatial epigenome–transcriptome mouse brain dataset, as well as $1.26\%$ and $2.08\%$ for the Spatial multi-modal omics simulation dataset.
In addition, for several common clustering evaluation indicators such as MI, AMI, FMI, ARI, V-Measure, Jaccard similarity, and Completeness, our method is also significantly better than the baseline methods.
Quantitative experimental results on the above three datasets demonstrate that our proposed PRAGA can obtain reliable comprehensive representations from spatial multi-modal omics data.

% \begin{figure*}
%   \centering
%   \includegraphics[width=1.0\textwidth]{Vis.pdf}
%   \caption{Visualization of qualitative experimental results on real datasets SPOTS and mouse thymus Stereo-CITE-seq.}
%   \label{_frame}
% \end{figure*}

\begin{table}[t]
\fontsize{9}{9}\selectfont
\centering
\setlength{\tabcolsep}{2.5mm}
\begin{tabular}{lcccccc}
\toprule
Init C & NMI & ARI  & F1-Score & Jaccard & Compl.\\
& (\%) & (\%)& (\%)& (\%)& (\%)\\
\midrule
5 & 38.48 & 27.09 & 41.21 & 25.95 & 35.19\\
6 & 38.59 & 26.71 & 40.51 & 25.40 & 35.23\\
7 & 39.00 & 27.32 & 41.21 & 25.95 & 36.02\\
8 & 37.01 & 24.19 & 40.51 & 25.40 & 33.34\\
9 & 38.07 & 25.84 & 39.31 & 24.47 & 34.46\\
10$\dag$ & \textbf{39.47} & \textbf{28.28} & \textbf{42.23} & \textbf{26.76} & \textbf{36.29} \\
11 & 38.00 & 26.02 & 39.47 & 24.58 & 34.36\\
12 & 38.84 & 26.79 & 39.49 & 24.40 & 34.89\\
13 & 39.06 & 27.63 & 41.45 & 26.14 & 35.81\\
14 & 37.95 & 26.62 & 40.20 & 25.16 & 34.52\\
15 & 38.68 & 27.62 & 42.02 & 26.60 & 35.75\\

\bottomrule
\end{tabular}
\caption{Quantitative performance of PRAGA with different initial cluster numbers.
$\dag$ marks the number of cluster categories in Ground Truth.
The best experimental results are marked in bold.}

\label{initk}

\end{table}

\subsection{Ablation studies}
In this subsection, we verify the effectiveness of the proposed learnable feature graph, homogeneity loss, reconstruction losses, and dynamic prototype contrastive learning loss on the Human Lymph Node Dataset.
As illustrated in Table~\ref{Ablation}, when the proposed learnable graph is replaced by the adjacency graph built by KNN, the performance of the PRAGA in MI, NMI, AMI, and Completeness decreases by $1.28\%$, $0.78\%$, $0.84\%$, and $0.79\%$ respectively.
We explain that the cause of this performance degradation phenomenon is the loss of latent correlations between sequencing points, which are well captured and integrated with spatial information through our proposed learnable graph.
The absence of homogeneity loss $\mathcal{L}_h$, reconstruction loss $\mathcal{L}_{recon}$, and dynamic prototype contrastive learning loss $\mathcal{L}_{dpcl}$ results in varying degrees of degradation in model performance, which in turn proves their performance contribution.

% The ablation results of the homogeneity loss $\mathcal{L}_f$ show that the homogeneity loss seems to contribute slightly to the performance of the model.
% However, the homogeneity loss prevents the learnable feature graph from being out of interpretability, $\textit{i.e.}$, the feature graph iteratively explores latent semantic correlations based on an initial KNN graph.

% The absence of reconstruction loss leads to a significant drop in model performance, which is attributed to the fact that the existence of reconstruction loss can protect the modality-specific information from being lost during the aggregation encoding process~\cite{long2024deciphering,li2024cross}.
% Removing our proposed dynamic prototype contrastive learning also has a negative impact on the performance of GRAGA. 
% This demonstrates that dynamic prototype contrastive learning can provide valuable cluster prototype knowledge to the model even in scenarios with unknown biological priors.

\begin{figure}
  \centering
  \includegraphics[width=0.5\textwidth]{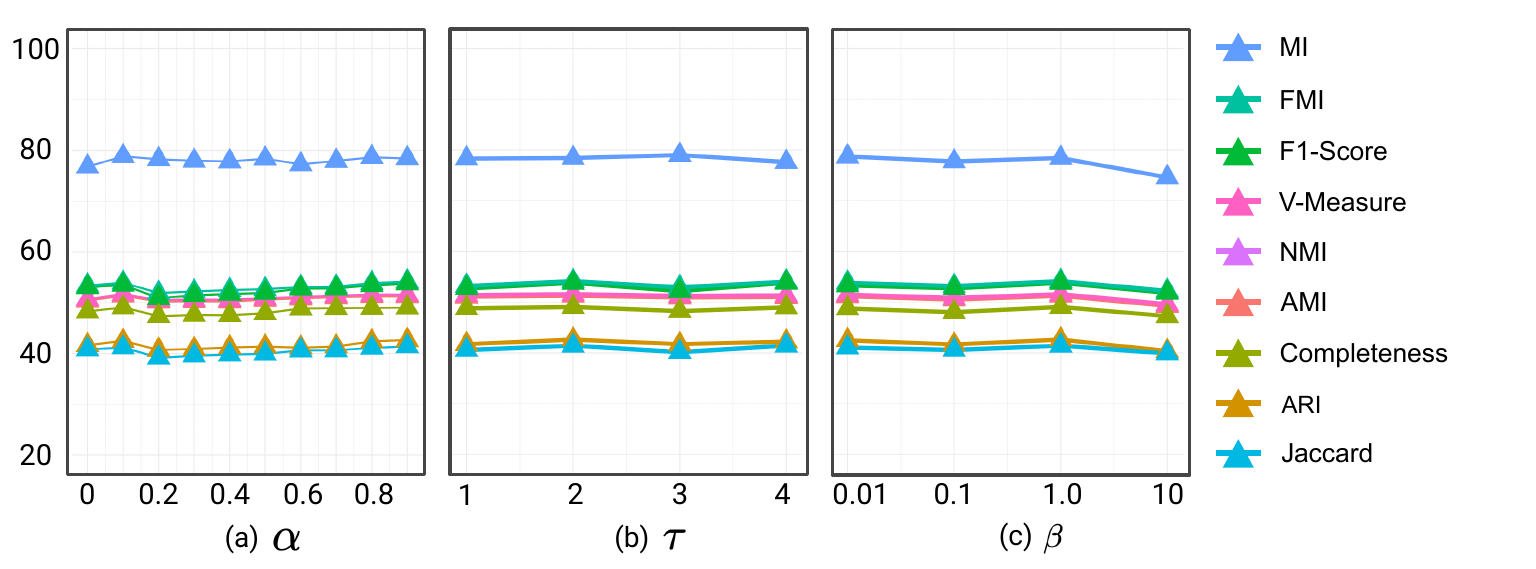}
  \caption{Effects of exponential moving average speed $\alpha$, temperature $\tau$ in $\mathcal{L}_{dpcl}$, and weight $\beta$ for $\mathcal{L}_{dpcl}$ on PRAGA performance measured by nine different metrics.}
  \label{pse}
\end{figure}

\subsection{Parameter sensitivity experiments}

We conduct parameter sensitivity experiments to verify the sensitivity of the performance of our proposed PRAGA to different values of hyperparameter, including the initial number of clusters $C$, exponential moving average speed $\alpha$, temperature hyperparameter $\tau$, and weight $\beta$ of $\mathcal{L}_{dpcl}$.
The experiments in this subsection are conducted on the Human Lymph Node dataset.

Table~\ref{initk} shows the quantitative performance of PRAGA with different initial cluster numbers.
Consistent with intuition, PRAGA performs best when the number of initial clusters is the same as the number of categories in Ground Truth.
Nevertheless, when the number of initial clusters differs from Ground Truth, which is also a common case in practice, PRAGA still performs well without significant performance degradation.
% This demonstrates that PRAGA is reliable even if the category priors of sequencing points are unknown.
Figure 4 shows the effect of different values of exponential moving average speed $\alpha$, temperature $\tau$ in $\mathcal{L}_{dpcl}$, and weight $\beta$ for $\mathcal{L}_{dpcl}$ on PRAGA performance.
Additional sensitivity experiments and the discussion of runtime can be found in the appendix.
The experimental results consistently show that PRAGA is insensitive to the values of hyperparameters.

\section{Conclusion}
In this paper, we propose a novel spatial multi-modal omics framework, named \textbf{PR}ototype-\textbf{A}ware \textbf{G}raph \textbf{A}daptative aggregation for spatial multi-modal omics analysis (PRAGA).
On the one hand, PRAGA performs the dynamic feature graph to denoise the sequencing perturbations by learning cross-modal semantics.
On the other hand, PRAGA integrates spatial information and multi-modal omic features to generate reliable comprehensive representations for downstream biological applications.
The dynamic prototypical contrastive learning is proposed to promote the dynamic feature graph to learn abundant latent semantic relations.
Qualitative and quantitative experimental results across 5 datasets demonstrate that our proposed PRAGA significantly outperforms existing State-Of-The-Art spatial multi-modal omics methods.

\section{Acknowledgements}
The project is supported by the National Natural Science Foundation of China (Grant No. 32300554 and No. 62406056), and in part by the Guangdong Provincial Key Laboratory of Mathematical and Neural Dynamical Systems (Grant No.2024B1212010004). The computational resources are supported by Songshan Lake HPC Center (SSL-HPC) at Great Bay University.

\bibliography{aaai25}

\end{document}